\documentclass[draftclsnofoot,12pt,onecolumn]{IEEEtran}
\usepackage{amsmath,amsfonts,amsthm,amssymb}
\usepackage{cite}
\usepackage{algorithm}
\usepackage{enumerate}
\usepackage{algpseudocode}
\usepackage{cases}
\usepackage{color}
\usepackage[dvips, final]{graphicx}

\newcommand{\figwidth}{.8\linewidth}
\normalsize

\def\figref#1{Figure~\ref{#1}}

\def\secref#1{Section~\ref{#1}}
\newcounter{remark}
\def\remark{\addtocounter{remark}1\noindent\emph{Remark \arabic{remark}:} }

\newcounter{testcase}
\def\testcase{\addtocounter{testcase}1\smallskip
\noindent\emph{Test Case \arabic{testcase}:} }

\newcounter{step}

\def\normal{f^{\text{(N)}}}

\DeclareMathOperator{\trace}{Tr}

\def\bA{{\mathbf A}}
\def\bB{{\mathbf B}}
\def\bC{{\mathbf C}}

\def\bI{{\mathbf I}}
\def\bP{{\mathbf P}}

\def\bY{{\mathbf Y}}
\def\bZ{{\mathbf Z}}
\def\bh{{\mathbf h}}
\def\br{{\mathbf r}}
\def\bs{{\mathbf s}}

\def\bv{{\mathbf v}}
\def\bx{{\mathbf x}}
\def\by{{\mathbf y}}
\def\bz{{\mathbf z}}

\def\tLambda{{\Tilde\Lambda}}

\def\bbs{{\Bar{\mathbf s}}}

\def\tbs{{\Tilde{\mathbf s}}}

\def\tbz{{\Tilde{\mathbf z}}}
\def\tbr{{\Tilde{\mathbf r}}}

\begin{document}

\setcounter{page}{1}

\title{Many Access for Small Packets Based on Precoding and Sparsity-aware
Recovery}

\author{
Ronggui Xie$^1$, Huarui Yin$^1$, Xiaohui Chen$^1$, and
Zhengdao Wang$^2$\\
\small $^1$Department of Electronic Engineering and Information Science,
University of Science and Technology of China\\ \small $^2$Department of
Electrical and Computer Engineering, Iowa State University}

\maketitle

\begin{abstract}
Modern mobile terminals produce massive small data packets. For these
short-length packets, it is inefficient to follow the current multiple
access schemes to allocate transmission resources due to heavy signaling
overhead. We propose a non-orthogonal many-access scheme that is well
suited for the future communication systems equipped with many receive
antennas. The system is modeled as having a block-sparsity pattern with
unknown sparsity level (i.e., unknown number of transmitted messages).
Block precoding is employed at each single-antenna transmitter to enable
the simultaneous transmissions of many users. The number of simultaneously
served active users is allowed to be even more than the number of receive
antennas. Sparsity-aware recovery is designed at the receiver for joint
user detection and symbol demodulation. To reduce the effects of channel
fading on signal recovery, normalized block orthogonal matching pursuit
(BOMP) algorithm is introduced, and based on its approximate performance
analysis, we develop interference cancellation based BOMP (ICBOMP)
algorithm. The ICBOMP performs error correction and detection in each
iteration of the normalized BOMP. Simulation results demonstrate the
effectiveness of the proposed scheme in small packet services, as well as
the advantages of ICBOMP in improving signal recovery accuracy and reducing
computational cost.

\end{abstract}

\begin{IEEEkeywords}
small packet, non-orthogonal many access, precoding, block-sparsity, error
correction and detection, order statistics
\end{IEEEkeywords}

\IEEEpeerreviewmaketitle

\section{Introduction} Intelligent terminals such as smart phones and tablets
are widely used nowadays, and their numbers are expected to grow in the
near future. It is likely that many applications in these intelligent
terminals are simultaneously requesting to communicate with a same base
station (BS). The resulting data packets to be delivered over a cellular
network are usually very short, i.e., small packets \cite{re2,re3}. Mobile
applications such as instant messaging are the main producers of small
packets. In the services of small packet, the number of users
simultaneously requesting services may even be comparable with the length
of coding block, which makes the channel a many-access channel
\cite{GuoDongning2013,GuoDongningarxiv}. On the other hand, the service
requests of small packet are frequent and irregular \cite{re4}. Although
many users may simultaneously have packets to send, they represent only a
small percentage of the total users connected to the network.

In current systems, small packet services put great burden on the cellular
network. One factor is due to the low efficiency of the transmitted packet.
Even very short conversation packet is accompanied by regular overhead such
as identity information. More importantly, for frequent and random
transmission requests, when following the existing multiple access schemes
that allocate orthogonal resources for different transmissions, signaling
overhead related to scheduling these packets is very heavy. For example,
even the scheduling-based schemes which are widely deployed in systems such
as long term evolution (LTE) to ensure the quality of services, will
produce massive overhead in dynamic resource programming. Future wireless
communication systems, e.g., 5G network \cite{pirinen2014brief}, are
required to offer a lot of such connections, which draws much attention
from academia and industry to propose flexible and efficient protocols,
e.g.,
\cite{GuoDongning2013,GuoDongningarxiv,zhang2014virtual,re8,SCMA2013,SCMAcodebook2014,NOMA2015,MUSA2015}
and references therein.

In this paper, for the services of small packet, we propose a non-orthogonal
many access scheme for uplink. The proposed scheme is based on precoding at
the transmitters and sparsity-aware recovery at the receiver. The main
motivation is to allow for a large number of users to transmit simultaneously,
although each user may transmit only a small amount of data. Block precoding
is employed to enable the simultaneous transmissions of many users. The
sparsity-aware recovery is designed to detect the user activity and recover
the transmitted data jointly. We say a user is \emph{active} if the user is
transmitting data. The user activity, including the number of active users and
the identities of active users, is not known at the BS. Besides frame-level
synchronization, no competition for resources or other coordination are
required. This saves the signaling overhead related to allocating resources
for transmission, and improves the resource utilization efficiency.

The contributions of our work are as follows:
\begin{enumerate}

\item \emph{Block precoding and block-sparse system modeling:} We apply
block precoding to short-length message at each transmitter, and by
considering the user activities, we develop a block-sparse system model
with unknown sparsity level. The established model takes full advantages of
the block transmission behavior and is suitable for sparse recovery based
algorithms.

\item \emph{Block-sparse recovery algorithms:} Signal recovery includes
user detection and the symbol demodulation in our scenario. To mitigate the
influences of channel state on recovery, we modify the existing BOMP
algorithm into a normalized BOMP algorithm. And then based on the
normalized BOMP, we develop an interference cancellation (IC) based BOMP
(ICBOMP) algorithm. The ICBOMP improves upon the normalized BOMP by taking
advantage of the availability of error correction and detection coding. By
the ICBOMP algorithm, we achieve better signal recovery accuracy and lower
computational cost. The price is a slightly decreased data rate due to
coding.

\item \emph{Signal recovery analysis:} We approximately analyze the
performance of signal recovery by the normalized BOMP algorithm.
Considering the characteristic of the sparse recovery, order
statistics \cite{ICanalysis1994} is used for the approximate
analysis. Among many mutually-exclusive cases of successful user detection,
we choose the most likely one to analyze its probability. The obtained
probability is used to serve as the approximate lower bound for the
probability of successful detection. Based on the successful detection, we
perform analysis for symbol error rate (SER).

\end{enumerate}

Thanks to the precoding operation and the designed sparsity-aware recovery,
our scheme enables the system to serve many active users simultaneously. The
number of active users can be even more than the number of antennas at the BS.
When BS is equipped with a large-scale antenna array \cite{re9,re10}, it can
offer small packet services to a very large number of users. At the same time,
as demonstrated by the sparse-recovery framework in \cite{Novelaccess2010},
our scheme can also provide benefits in saving the identity information and
reducing the decoding delay. Therefore, the proposed scheme is especially
suitable for the future 5G system.

The most related works to our proposed scheme are sparsity-aware recovery
based multi-user detection studied in papers such as
\cite{GuoDongning2013,GuoDongningarxiv,zhang2014virtual,Exploitingsparse2011,Reduced2013,OnOffCS,CSMUDMtM2013,Novelaccess2010,MPACSJung2013,CSmeterreading2010,Guo2011,DistributedaccessCS2012}
and references therein. In most of the referred works, sparse transmissions
have been considered for the detection task. Some of the works address the
problem of user detection and the afterwards symbol demodulation. A new
notion of capacity related to many-access channel is introduced in
\cite{GuoDongning2013,GuoDongningarxiv}. Full duplex is virtually realized
in \cite{zhang2014virtual}, where on and off slots are respectively
assigned for transmitting to and receiving from neighbors, and finally
sparse recovery is done at the receiver when one frame transmission completes.
In \cite{Exploitingsparse2011}, based on convex optimization, the authors
introduce sparsity-aware detectors for the multi-user detection in
code-division multiple access (CDMA) systems. The fact that sparse recovery
algorithms such as OMP can provide better performance than single-user
detection is demonstrated in \cite{OnOffCS}. Works in
\cite{Novelaccess2010,CSmeterreading2010,DistributedaccessCS2012} show
that, compared with some well known schemes, the CS based multiple access
scheme can save the identity information overhead and reduce the decoding
delay. A CS based multi-user detection scheme for machine-to-machine (M2M)
communication can be found in \cite{CSMUDMtM2013}. Particularly, authors in
\cite{CSMUDMtM2013} establish a block-sparse system model and adopt a novel
coding scheme to improve the user detection accuracy.

Our work differs from the above referred works. All works in
\cite{GuoDongning2013,GuoDongningarxiv,zhang2014virtual,Exploitingsparse2011,Reduced2013,OnOffCS,CSMUDMtM2013,Novelaccess2010,CSmeterreading2010,Guo2011}
are about scenario of single-antenna receiver. Works in
\cite{Reduced2013,OnOffCS,Novelaccess2010,MPACSJung2013,CSmeterreading2010,Guo2011,DistributedaccessCS2012}
establish sparsity model where nonzero coefficients to detect are just
randomly located among all possible positions in a sparse vector, which
neglects the fact that the transmission of a message could last for a few
symbol slots, i.e., block transmission. Block transmission is considered in
\cite{zhang2014virtual,CSMUDMtM2013}. BOMP algorithm is adopted in
\cite{CSMUDMtM2013} for the data recovery and an activity-aware channel coding
is used to improve the recovery performance, but coding scheme in it is not
fully incorporated into the greedy algorithm to reduce the computational cost.

The rest of the paper is organized as follows. In Section
\ref{SystemModel}, the system model of block sparsity is given. In Section
\ref{BOMP}, we introduce the normalized BOMP algorithm to recover the
transmitted signals and analyze its performance. Improved algorithm version
ICBOMP is proposed in Section \ref{ICBOMP}. Section \ref{simulations}
presents the numerical simulation results that demonstrate the
effectiveness of the proposed scheme. Some issues are discussed in Section
\ref{Discussion}. Finally, conclusions are presented in Section
\ref{Conclusion}.

Notation: Vectors and matrices are denoted by boldface lowercase and uppercase
letters, respectively. The 2-norm of a vector $\bv$ is denoted as ${\| \bv
\|_2}$, and $\bv^T$ denotes its transpose. The identity matrix of size $n
\times n$ is denoted as $\bI_n$. For a matrix $\bA$, its conjugate transpose
is denoted as $\bA^H$. The $(i,j)$-th entry of a matrix $\bA$ is denoted as
$[\bA]_{i,j}$. Operation $vec(\bA)$ denotes vectorization of $\bA$ by column
stacking. For a subset $I\subset [N]:=\{1,2, \cdots ,N\}$ and matrix $\bA: = [
{{\bA_1},{\bA_2}, \cdots ,{\bA_N}}]$ consisting of $N$ sub-matrices (blocks),
where each sub-matrix has an equal size, $\bA_I$ stands for a sub-matrix of
$\bA$ whose block indices are in $I$; for a vector $\bv: = {[
{\bv_1^T,\bv_2^T, \cdots ,\bv_N^T} ]^T}$, $\bv_I$ is similarly defined. Value
$|I|$ stands for the cardinality of set $I$. Given two sets $I_1$ and $I_2$,
$I_1\backslash I_2:=I_1\cap I_2^c$. For a real number $r$, $\lfloor r \rfloor
$ and $\left\lceil r \right\rceil $ respectively stand for its floor and
ceiling. Operators $\otimes$, $\trace$ and $\mathbb{E}$ respectively stand for
the Kronecker product, the trace of a matrix and the expectation of a random
variable. $Re$ stands for taking the real part from a complex number, and
$\binom{m}{k}$ is the binomial coefficient for $m\ge k$. The real Gaussian PDF
with mean $\mu$ and variance $\sigma^2$ is denoted as
\begin{equation}
f^\text{(N)}(x|\mu,\sigma^2)=\frac 1{\sqrt{2\pi\sigma^2}}
\exp\left(-\frac{(x-\mu)^2}{2\sigma^2}\right).
\end{equation}

\section{System Model}\label{SystemModel}

Consider an uplink system with $N$ mobile users, each with a single
antenna, and a BS with $M$ antennas. When a terminal is admitted to the
network, it becomes an \emph{online} user. We assume that there are
$N_\text{a}$ \emph{active} users, out of the total $N$ online users, that
simultaneously have data to transmit. It is not required $N_\text{a}$ be
known \emph{a priori} or ${N_\text{a}} < M$. Actually in practical systems,
$N_\text{a}$ is usually unknown and it is possible that $N_\text{a} \ge M$.

We make the following additional assumptions on the system considered.
\begin{enumerate}
\item The channels are block-fading. They remain constant for a certain
duration and then change independently.

\item The transmissions are in blocks and the users are synchronized at the
block level. We assume that each frame of transmission consists of $T$
symbols, which all fall within one channel coherent interval. Value $T$
depends on the communication standard.

\item The antennas at the BS, as well as the antennas among users, are
sufficiently apart to yield spatially independent channels.

\item The BS always has perfect channel state information (CSI) of online
users.
\end{enumerate}

Since the lengths of small packets are usually shorter than $T$, we extend
their lengths by precoding to fully utilize the available resources. Let
$\bs_n \in {\mathbb{C}^{d \times 1}}$ denote the symbol vector to be
transmitted by user $n$, with $d < T$, $n=1,2,\cdots,N$. User $n$ applies a
precoding to $\bs_n$ to yield
\begin{equation}
{\bx_n} = {\bP_n}{\bs_n}
\end{equation}
where $\bP_n$ is a complex precoding matrix of size $T\times d$. We assume
that $\bP_n$ is normalized such that each column has 2-norm equal to 1. The
entries of $\bx_n$ are transmitted in $T$ successive time slots.\footnote{ In
fact, $T$ can be generally regarded as the number of resource units in time,
frequency and code domains. For example, $T$ resource units can be composed of
$m$ subcarriers and $\lceil T/m \rceil$ successive time slots for each
subcarrier.} The received signals at the receiver within one frame can be
written as
\begin{equation}
\bY = \sqrt {{\rho_0}} \sum_{n = 1}^N {{\bh_n}\bx_n^T} +\bZ
=\sqrt {{\rho_0}} \sum_{n = 1}^N {{\bh_n}\bs_n^T{\bP_n^T}} +\bZ
\end{equation}
where ${\rho_0}$ is the signal to noise ratio (SNR) of the uplink, $\bY$ is
the noisy measurement of size $M\times T$, $\bZ \in {\mathbb{C}^{M \times T}}$
represents the additive Gaussian noise matrix of complex-valued, and $\bh_n
\in {\mathbb{C}^{M \times 1}}$ represents the complex channel response from
user $n$ to the BS. Without loss of generality, each element of $\bh_n$,
$\bZ$ and nonzero $\bs_n$ is assumed to have a zero mean and a unit variance. Using the linear
algebra identity $vec({\bA\bB\bC}) = ({{\bC^T} \otimes \bA})vec(\bB)$, we can
rewrite the received signal as
\begin{equation} \label{model0}
vec(\bY) =\sqrt {{\rho_0}} \sum_{n = 1}^N {(\bP_n \otimes {\bh_n}){\bs_n}}
 +vec(\bZ).
\end{equation}

Define $\by: = vec(\bY)$, ${\bB_n}: = {\bP_n \otimes {\bh_n}} $, $\bB: = [
{{\bB_1},{\bB_2},\cdots,{\bB_N}} ]$ and $\bs: = {[
{\bs_1^T,\bs_2^T,\cdots,\bs_N^T} ]^T}$. Then \eqref{model0} can be
rewritten as
\begin{equation} \label{model1}
{\by} =\sqrt {{\rho_0}} {\bB}{\bs}+ \bz.
\end{equation}

In the above formulation, we have assumed that all messages have an equal
length $d$. We view $d$ as the maximum length of the messages for all users
within a frame. For the users whose message lengths are less than $d$, we
assume their messages have been zero-padded to $d$ before precoding. For
those users that are not active, we view their transmitted symbols as all
zeros. Throughout this paper, we say $\bs_n$ is a block of $\bs$ and
$\bB_n$ is a block of $\bB$.

Since typically only a small percentage of the online users are active,
nonzero signals are only located in a small fraction of blocks in $\bs$ and
all other blocks are zero. When $MT > Nd$, the receiver design is easy. Our
consideration is limited to case $MT < Nd$, where \eqref{model1} has the
same form as an overloaded CDMA system. The equivalent spreading chip
sequences in $\bB$ no longer have constant amplitude and are generated by
combining precoding codes and channel gains. When precoding
matrix ${\bP_n}$ is well designed, matrix ${\bB}$ can meet the restricted
isometry property (RIP) requirement for measurement matrix in the CS literature
\cite{re13,re16}. Therefore, from the viewpoint of signal recovery,
\eqref{model1} can be viewed as a block-sparsity CS model \cite{re24}.

Although the information of active users is unknown at the BS, it is assumed
that the BS knows the sparsity of transmission, so that it can perform
sparsity-aware recovery. Let $I$ be the set containing the unknown indices
of active users, with $| I | = N_\text{a} \le N_{\text{a}\max}$. This means
that the unknown number of active users is at most $N_{\text{a}\max}$. We
assume $N_{\text{a}\max}$ is known at the receiver.

\remark The precoding scheme is proposed because in reality, $T$ is usually
longer than the lengths of small packets. Also, the precoding scheme
contributes to solving the signal recovery problem in the situation where
$N_\text{a}>M$.

\remark Each user knows its own precoding matrix and the BS knows all users'
precoding matrices. Since any two precoding matrices are not allowed to be
identical, each precoding matrix represents a unique user. Therefore, no
additional identity information of users is required.

\remark The precoding matrix should have full column rank for data recovery.
Additionally, we assume each column of the precoding matrices is normalized
to have unit energy.

\section{Normalized BOMP Algorithm For Signal Recovery}\label{BOMP}

Sparsity-aware recovery has been widely studied in the frame work of CS
\cite{re13,re16}. Initial work in CS treats sparse weighting coefficients
as just randomly located among all possible positions in a data vector,
i.e., random sparsity. When block-sparsity of the vector is taken into
account, it is possible to obtain better recovery performance and reduce
the number of required measurements \cite{re20,re24,re25,kwon2012benefits}.

As for signal recovery algorithm to be used at the BS, we first apply the
known BOMP algorithm and modify it to our problem. The normalized BOMP
algorithm is therefore introduced. Then, we analyze the performance of the
normalized BOMP, including user detection and signal demodulation. The
order statistics \cite{ICanalysis1994} is essential for our analysis.

\subsection{Normalized BOMP algorithm} The main idea of BOMP is that, at each
iteration, it chooses one block from $\bB$ which has the maximum
correlation with the residual signal. After that, it will use all the
previously selected blocks to update the signals by solving a least-square
(LS) problem \cite{re24}. Due to ${\bB_n} = {\bP_n \otimes {\bh_n}} $ and
to mitigate the negative influence of $\bh_n$ on user detection, it is
beneficial to use the norm of $\bh_n$ to normalize the columns of ${\bB_n}$
when calculating the correlation value. Idea of similar modification can
also be found in \cite{CSMUDMtM2013} to combat the influences of channels
in a CDMA system. Detailed iterations of normalized BOMP is presented in the
Algorithm 1, where $c_{j, k}={\| {{\bB_j^H}{\br_{k-1}}}
\|_2^2/\|\bh_j\|_2^2}$ is the squared correlation coefficient of normalized.

\alglanguage{pseudocode}
\begin{algorithm}
\caption{normalized BOMP} \label{algorithm1}
\begin{algorithmic}[1]
\State \textbf{Input: }${\bB}$, $\by$, ${\rho_0}$, all channel vectors,
iterations $K$ (necessarily $K \le \lfloor \frac{MT}d \rfloor $ or directly
$K = N_{\text{a}\max}$) \State \textbf{Initialization: }index set
${{\Lambda}_0} = \phi $, residual signal ${\br_0} = \by$ \ForAll{$k \in
\{1, 2, \cdots, K\}$}
      \begin{enumerate}

      \item calculate the $c_{j, k}$ for $j \in \left\{[N]\backslash {{\Lambda}_{k-1}}\right\}$: $c_{j, k}={\| {{\bB_j^H}{\br_{k-1}}} \|_2^2/\|\bh_j\|_2^2}$
      \item find the index ${\lambda_k}$:
            $ {\lambda_k} = \arg \max_{j \in  \left\{[N]\backslash {{\Lambda}_{k-1}}\right\}} c_{j, k}$
      \item augment the index set:
       ${{\Lambda}_k}= {{\Lambda}_{k-1}} \cup \{ {\lambda_k}\}$
      \item update the signals by LS algorithm:
           $
              {\bbs_{{\Lambda}_k}} = \arg \min_{{\bs_0}}
              {\| {\by-\sqrt {{\rho_0}} {\bB_{\Lambda_k}}{\bs_0}}\|_2} $

      \item update the residual signals:
           $
              {\br_k} = \by-\sqrt {{\rho_0}}{\bB_{\Lambda_k}}{\bbs_{{\Lambda}_k}} $

      \end{enumerate}
\EndFor \State \textbf{Output: }${\bbs_K}$: the reconstruction of most
likely signal-bearing blocks in the index set ${\Lambda_K}$
\end{algorithmic}
\end{algorithm}

With the reconstructed coefficients, we will then demodulate the
transmitted messages.

\subsection{Analysis of group user detection probability}

Since the identities of the active users are unknown to the BS, the BS need to
first perform detection of the active users' identities before symbol
demodulation can be performed. At the $k$-th iteration, in order to guarantee
that an inactive user is not falsely added to the list of users to be
demodulated, we need $\max_{j\in I} c_{j, k} > \max_{j\notin I} c_{j, k}$. In
our analysis below, we will use group user detection success rate (GUDSR) as
the performance metric, which is defined as the probability that all active
users are selected by the recovery algorithm, i.e., $I\subset {\Lambda}_K$.
When $K > N_\text{a}$, as many as $K-N_\text{a}$ non-active users will be
falsely selected as active, this is unavoidable for our normalized BOMP
algorithm because the number $N_\text{a}$ of active users is unknown and no other
side information is available or utilized to decide whether a user is actually
active. The situation can be greatly improved when error control coding is
used (see \secref{ICBOMP}).

Exact active user detection probability analysis is difficult for several
reasons, including 1) the distribution of the entry of $\bB$ is difficult to
characterize because it involves both the linear precoding code and the
channel fading, and 2) error propagation in the normalized BOMP iterations
cannot be avoided. To make the analysis tractable, we make the following
approximations:
\begin{enumerate}
\item We assume that the entries of precoding matrices are i.i.d.\ complex
Gaussian variables, with zero mean and variance $\frac{1}{T}$. When $d$ and
$T$ are large enough, such approximation is reasonable for the analysis of the
statistics of the correlation coefficients.

\item Although the normalized BOMP algorithm works with the channel
realizations, we evaluate the performance by averaging over channel fading
statistics, which makes the analysis tractable. The result will therefore
depend on the channel statistics rather than specific channel realizations.

\item We approximate the distributions of the $c_{j, k}$ for both $j\notin I$
and $j\in I$ as Gaussian. Even though by definition $c_{j,k}$ is non-negative,
the approximation is reasonable when $dT$ is large.
\end{enumerate}
In the following, we will analyze the performance of user detection by several
steps. The order of selecting active users is firstly studied and then order
statistics is applied for the analysis.

\subsubsection{User selection order}

At the start of the $k$-th iteration, the residual signal from the previous
iteration is
\begin{align}
\br_{k-1}  &= \by - \sqrt {\rho_0} \bB_{ {\Lambda_{k-1}}} \bbs_{{{\Lambda}_{k-1}}} \label{residualt0a} \\
&= \by - \sqrt {\rho_0} \bB_{ {\Lambda_{k-1}}} \left(\bB_{{\Lambda_{k-1}}}^H \bB_{ {\Lambda_{k-1}}} \right)^{-1}
\bB_{{\Lambda_{k-1}}}^H \by \label{residualt0b}\\
&=  \sqrt{\rho_0}\sum_{i \in I \backslash {\Lambda_{k-1}}}\bB_i \bs_i + \tbz_k  \label{residualt0c}
\end{align}
where $\tbz_k = \bz -\bB_{ {\Lambda_{k-1}}}\left(\bB_{{\Lambda_{k-1}}}^H \bB_{
{\Lambda_{k-1}}} \right)^{-1} \bB_{{\Lambda_{k-1}}}^H \left( \sqrt{\rho_0}
\sum_{i \in I \backslash {\Lambda_{k-1}}} \bB_i \bs_i + \bz \right)$, and the
second term of $\tbz_k$ is the propagating error of the normalized BOMP
resulting from previous iterations.
We will first evaluate the means of the
squared correlation coefficients $c_{j,k}$ between the residual signal
$\br_{k-1}$ and the normalized signature sequences $\frac 1{\|\bh_j\|_2}
\bB_{j}$ for both active and inactive users. Let $\mu_{0, k, j}$ and $\mu_{1,
k, j}$ denote the mean values of $c_{j, k}$ for $j \notin I$ (inactive user)
and $j\in I$ (active user), respectively. Let $s_k$ be the number of active
users that have been selected by the first $k$ iterations. Let $\sigma_k^2 $
denote the variance of each element of $\tbz_k$. Then we have
\begin{align}
\mu_{0, k, j} & \approx (N_\text{a}-s_{k-1})\frac{\rho_0 d^2}{T} + d
  \sigma_k^2, \label{mean00}\\
\mu_{1, k, j} & \approx \frac{\rho_0 d}{T} (d+T-1) \mathbb{E} \left\{\bh_j^H
  \bh_j \right\} + \rho_0 (N_\text{a}-s_{k-1}-1)\frac{d^2}{T} + d \sigma_k^2.
  \label{mean01}
\end{align}
Results in \eqref{mean00} and \eqref{mean01} have been obtained by averaging
over all channel realizations. See Appendix A for the derivations of
\eqref{mean00} and \eqref{mean01} and the approximations involved. Generally
speaking, among the remaining active users that have not been selected, the one
with the largest $\bh_j^H \bh_j$ is expected to be selected at each iteration.
This can also be observed from \eqref{mean01} --- if the channels are random,
then the one with the largest energy $\mathbb{E} \left\{\bh_j^H \bh_j
\right\}$ will result in the largest mean in the squared correlation
coefficient. Next, we will apply order statistics to derive the distributions
of $ \bh_j^H \bh_j $ for different active users.

\subsubsection{Order statistics of channel coefficients}

Without loss of generality, we assume $I=[N_\text{a}]$ and $\bh_1^H \bh_1 \ge
\bh_2^H \bh_2 \ge \cdots \ge \bh_{N_\text{a}}^H \bh_{N_\text{a}}$. Indices
from $N_\text{a}+1$ to $N$ are for the inactive users. For each user $j$, the
\emph{unordered} random variable $\bh_j^H \bh_j$ follows a chi-squared
distribution with $2M$ degrees of freedom. The PDF $f(x)$ and CDF $F(x)$ of
the unordered $\bh_j^H \bh_j$ are given as follows:
\begin{equation} \label{pdf}
f(x) ={\exp\left( - x \right)} \frac{{x}^{M-1}}{ (M-1)! }
\end{equation}
\begin{equation} \label{Cpdf}
F(x) =1-{\exp\left( - x \right)} \sum_{k=0}^{M-1}\frac{{x}^k}{ k! }
\end{equation}
both of which hold for $x\ge 0$, and are zero when $x < 0$. By order
statistics \cite{ICanalysis1994}, the PDF of the \emph{ordered} $\bh_{n}^H \bh_{n}, n
= 1, 2, \cdots, N_\text{a}$, is given as
\begin{equation} \label{Order}
f_n(x)=\frac{N_\text{a} !}{(N_\text{a}-n)!(n-1)!} F^{N_\text{a}-n}(x)[1-F(x)]^{n-1} f(x)
\end{equation}
for $x\ge 0$. The mean $\mathbb{E} \left\{\bh_{n}^H \bh_{n}\right\}$ can then be computed based on
the above PDF.

\subsubsection{Variance of residual error}

With $K$ iterations, there are many cases for a successful group user
detection. To simplify the study of GUDSR, we consider only one case for
simplicity: the first $N_\text{a}$ iterations select all the $N_\text{a}$
active users and the selection order is based on the descending order of
their $\bh_n^H \bh_n$. Since there are other cases where the active users can be
correctly identified, the GUDSR we obtain by considering only one case of the
successful detection will serve as a lower bound to the true GUDSR. The
assumed selection order is in line with a similar idea used to analyze the
performance of successive IC in \cite{ICanalysis1994}. Next, we will analyze
the probability of the occurrence of the considered case. Our analysis of
GUDSR will just involves the first $N_\text{a}$ iterations. At the $k$-th
iteration, the algorithm selects the $k$-th active user, i.e., the active user
with $\bh_k^H \bh_k$.

When the above selection order is considered in the analysis, for $1\le k \le
N_\text{a}$ and after the previous $k-1$ iterations, the variance $\sigma_k^2$
of $\tbz_k$ (c.f.~\eqref{residualt0c}) is given by
\begin{align} \label{sigmak2}
\sigma_k^2 &= \frac{1}{MT}\mathbb{E} \left\{\tbz_k^H \tbz_k \right\} \\
&=  \frac{1}{MT} \mathbb{E} \left\{ \bz^H \left[ \bI_{MT} - \bB_{\Lambda_{k-1}}\left(\bB_{\Lambda_{k-1}}^H \bB_{\Lambda_{k-1}} \right)^{-1}
\bB_{\Lambda_{k-1}}^H  \right]\bz \right\} \\
&\quad +   \frac{\rho_0}{MT} \sum_{n=k}^{N_\text{a}} \mathbb{E} \left\{ \bs_n^H \bB_n^H \bB_{\Lambda_{k-1}}\left(\bB_{\Lambda_{k-1}}^H \bB_{\Lambda_{k-1}} \right)^{-1} \bB_{\Lambda_{k-1}}^H \bB_n \bs_n \right\} \\
&\approx \frac{1}{MT} \trace \left\{\mathbb{E} \left\{ \bI_{MT} - \bB_{\Lambda_{k-1}}\left(\bB_{\Lambda_{k-1}}^H \bB_{\Lambda_{k-1}} \right)^{-1}
\bB_{\Lambda_{k-1}}^H  \right\} \right\}\\
&\quad + \frac{\rho_0 d}{M^2T^2}  \trace \left\{ \mathbb{E} \left\{ \bB_{\Lambda_{k-1}}\left(\bB_{\Lambda_{k-1}}^H \bB_{\Lambda_{k-1}} \right)^{-1} \bB_{\Lambda_{k-1}}^H \right\} \right\}  \sum_{n=k}^{N_\text{a}} \mathbb{E} \left\{\bh_n^H \bh_n \right\} \\
&= 1-\frac{(k-1)d}{MT} + \frac{\rho_0 (k-1) d^2}{M^2T^2}\sum_{n=k}^{N_\text{a}} \mathbb{E} \left\{\bh_n^H \bh_n \right\}
\end{align}
where the identity $\bB_n^H \bB_n=\left( {\bP_n^H \otimes \bh_n^H} \right) \left( {\bP_n
\otimes \bh_n} \right) = \left(\bP_n^H \bP_n\right) \otimes \left(\bh_n^H
\bh_n\right) =\left(\bP_n^H \bP_n\right) \left(\bh_n^H \bh_n\right)$ has been
used in the above derivation.

\subsubsection{Statistics of the squared correlation}

At the $k$-th iteration, by our previous assumption on the active user
selection order, the algorithm will select the $k$-th active user ranked
according to channel strengths. Since the statistics of the squared
correlation coefficient $c_{j,k}$ for inactive user $j\notin I$ does not depend
on the inactive user index, we use $\mu_{0 ,k}$ and $\sigma_{0 ,k}^2$ to
denote the mean and variance for $c_{j, k}$ for $j \notin I$. For $c_{k,k}$,
which corresponds to the $k$-th active user, we denote its mean and variance
as $\mu_{1 ,k}$ and $\sigma_{1 ,k}^2$, respectively. We have the following
results\footnote{When $k=N_\text{a}$, the $\sigma_{0 ,k}^2$ can be more
precisely given by using the result of \eqref{var002}
as the secondary moment.}
\begin{align}
\mu_{0 ,k} & \approx \frac{\rho_0d^2}{MT} \sum_{n=k}^{N_\text{a}} \mathbb{E}
\left\{\bh_n^H \bh_n \right\} + d \sigma_k^2 \label{mean0} \\
\sigma_{0 ,k}^2 &\approx d(d+1)\sigma_{\br_{k-1}}^4 - \mu_{0 ,k}^2
\label{var0} \\
\mu_{1 ,k} & \approx \frac{\rho_0 d}{T} (d+T-1) \mathbb{E} \left\{\bh_{k}^H
\bh_{k} \right\}
 + \frac{\rho_0d^2}{MT} \sum_{n=k+1}^{N_\text{a}} \mathbb{E} \left\{\bh_n^H \bh_n \right\} + d \sigma_k^2   \label{mean1} \\
\sigma_{1 ,k}^2 &\approx \frac{\rho_0^2}{T^2} (d^2+d)(d+T-1)^2 \mathbb{E}
\left\{\left(\bh_{k}^H \bh_{k} \right)^2 \right\} + d(d+1)\sigma_{\tbr_k}^4 \\
& \quad +\rho_0 d \left[ \frac{3}{4}+ 2d + \frac{2d^2}{T} + \frac{d}{4T} -
\frac{d^2}{2 T^2} \right] \sigma_{\tbr_k}^2 \mathbb{E} \left\{\bh_{k}^H
\bh_{k} \right\} - \mu_{1 ,k}^2
\label{var1}
\end{align}
where $\sigma_{\br_{k-1}}^2$ is the variance of each elements of the residual signal $\br_{k-1}$, given by
\begin{align}
\sigma_{\br_{k-1}}^2  &= \frac{1}{MT}\mathbb E \left\{ \br_{k-1}^H \br_{k-1}  \right\} \label{residulcovariance} \\
&= \frac{\rho_0}{MT} \sum_{n \in I \backslash {\Lambda_{k-1}} } \bs_n^H \bB_n^H \bB_n \bs_n   + \frac{1}{MT} \tbz_k^H \tbz_k \label{residulcovarianceb} \\
&\approx  \frac{\rho_0 d}{MT} \sum_{n=k}^{N_\text{a}} \mathbb{E} \left\{\bh_n^H \bh_n \right\} + \sigma_k^2  \label{residulcovariancec}
\end{align}
and $\sigma_{\tbr_k}^2$ is the variance of each elements of $\tbr_{k}: =\br_{k-1}-\sqrt{\rho_0} \bB_k \bs_k$, given as
\begin{equation}
\sigma_{\tbr_k}^2 \approx \frac{\rho_0
d}{MT} \sum_{n=k+1}^{N_\text{a}} \mathbb{E} \left\{\bh_n^H \bh_n \right\} +
\sigma_k^2,
\end{equation}
which is reduced to $\sigma_{\tbr_k}^2= \sigma_k^2$ when $k=N_\text{a}$.
The derivations of \eqref{mean0}-\eqref{var1} are presented in the Appendix B.
With the Gaussian assumption on $c_{j, k}$ in general and $c_{k,k}$ in
particular, the PDF of $c_{k,k}$ for the $k$-th active user is $f_{1, k} (x) =
\normal(x | \mu_{1,k}, \sigma_{1,k}^2)$. The PDF of
$c_{j,k}$ for a generic inactive user $j\not\in I$ is given by $f_{0, k} (x) =
\normal(x | \mu_{0,k}, \sigma_{0,k}^2)$, which does not depend on the index of
the inactive user.

\subsubsection{Successful user detection probability}

False alarm in the detection is generated in the first $N_a$ iterations of the
normalized BOMP algorithm when an inactive user has a larger squared correlation
$c_{j,k}$ than an active user. Using order statistics, the PDF of the
$\max_{j\notin I} c_{j, k}$ is given by
\begin{equation}   \label{maxgaussian0}
f_{0, k, \text{max}}(x) = (N-N_\text{a}) [ F_{0, k} (x) ]^{N-N_\text{a}-1} f_{0, k} (x)
\end{equation}
where $F_{0, k} (x)$ is the CDF of $c_{j, k}$ for $j\not\in I$.

Let $E_k$ denote the event that the $k$-th active user is selected at the
$k$-th iteration, i.e., $c_{k,k} > \max_{j\notin I} c_{j, k}$, whose probability is given
by
\begin{equation}   \label{prot1}
P (E_k) = \int_{-\infty}^{+\infty} \left[\int_{-\infty}^{x_1} f_{0, k, \text{max}}(x_0) d x_0 \right] f_{1, k} (x_1) d x_1,
\end{equation}
The successful group user detection of the considered decoding order has a
probability as follows
\begin{equation}   \label{protconsidered}
P^{(\text{u})} = \prod_{k=1}^{N_\text{a}}
  P \left( E_k | E_1,E_2,\cdots,E_{k-1}\right)
  \approx \prod_{k=1}^{N_\text{a}} P(E_k)
\end{equation}
where the last approximation is valid when the probability of $E_1,\ldots,
E_{k-1}$ is high, namely when the correct detection probability is high.

\remark Due to the approximations made in the derivation of $P^{\text{(u)}}$,
it can be used as an approximate lower bound of the true GUDSR only. In the simulations section,
we will demonstrate the usefulness of the derived lower bound using numerical
examples, when the SNR is not too small.

\remark The above analysis about GUDSR shows that, large $N_\text{a}$ and $N$
are not good for user detection when other parameters are given. It is because
they make the gap between normalized correlation coefficients of the last
active user and non-active user small, which is quantified by the order
statistics.

\remark Probability of successful detection of \emph{individual} active users
is another useful performance metric. However it is more difficult to analyze.
We will resort to numerical simulations for their evaluations. The error rate
of group detection is naturally typically higher than that of individual
detection because any individual detection error causes the group detection
failure.

\subsection{Symbol error rate analysis}

For the analysis of symbol demodulation error probability, we assume that
quadrature phase shift keying (QPSK) modulation is used as an example. Other
modulation schemes can be similarly considered. 

When all the $N_\text{a}$ active users and $K-N_\text{a}$ non-active users
have been selected by $K$ iterations of normalized BOMP, signal update is
finally as follows
\begin{equation} \label{updateK}
\bbs_{\Lambda_K}=\bs_{\Lambda_K} + \frac{1}{\sqrt {\rho_0}} \left( \bB_{\Lambda_K}^{H}\bB_{\Lambda_K}\right)^{-1}\bB_{\Lambda_K}^{H}\bz
\end{equation}
where $\bB_{\Lambda_K}$ is $MT \times Kd$ matrix composed of $K$ selected
blocks.

When the entries of $\bB_{\Lambda_K}$ are i.i.d.\ complex Gaussian variables,
the demodulation performance of \eqref{updateK} has been analyzed through an
upper bound in \cite{winters1994impact}. Our model is different in several
aspects: 1) each column of $\bB_{\Lambda_K}$ is a Kronecker product of a
precoding vector and a single-input multiple-output channel response vector;
2) every $d$ columns of $\bB_{\Lambda_K}$ share the same channel vector; 3)
the precoding vectors are normalized. Despite these differences, we can obtain
an approximated SER expression by assuming that the entries of
$\bB_{\Lambda_K}$ are i.i.d.\ complex Gaussian variables with zero mean and
variance $1/T$. Under this assumption, all symbols of all active users have
the same average (ergodic) performance, where the average is over channel
fading and noise statistics. As a result we can focus on the analysis on any
one symbol, say the first symbol of the first (unordered) active user.

Assuming without loss of generality that the first column of $\bB_{\Lambda_K}$
belongs to an active user. Let $G = 1/\left[\left( \bB_{\Lambda_K}^{H}
\bB_{\Lambda_K}\right)^{-1}\right]_{1,1}$, which is a random variable that
affects the SNR of the useful symbol. Let $\mathbf{b}_{1}$ be the first column
of $\bB_{\Lambda_K}$. Using the result in
\cite[Appendix~A]{winters1994impact}, we can express $G$ as
\begin{equation}  \label{impact1994a}
G = \sum_{m=1}^{MT-Kd+1} |b_m|^2
\end{equation}
where $b_m$, $m=1,\ldots, MT-KT+1$, are i.i.d.\ and they have the same
distribution as the entries of $\mathbf b_1$, namely zero-mean, circularly
symmetric complex Gaussian, with variance $1/T$. As a result, $G$ is a
scaled chi-squared distributed random variable with degrees of freedom
$2(MT-Kd+1)$, with PDF given as follows:
\begin{equation}
  f_G(g) = \frac{T^{MT-Kd+1}}{(MT-Kd)!} \cdot e^{-Tg} g^{MT-Kd} .
\end{equation}
Conditioned on $G=g$, the SER of the symbol of interest is given by
\begin{equation}   \label{SERn0}
\text{SER}(g) = \text{erfc} \left( \sqrt {0.5 \rho_0 g}\right) -
\left[ \frac{1}{2} \text{erfc} \left( \sqrt {0.5 \rho_0 g}\right) \right]^2.
\end{equation}
The average SER can then be evaluated as
\begin{equation}
\text{SER}=\int_0^\infty
\text{SER}(g)\cdot f_G(g) dg.
\end{equation}

\remark When $MT-Kd+1 \gg 1$, $G$ is well approximated by a constant
\begin{equation}\label{eq.G0}
G_0=\frac{MT-Kd+1}{T}= M- \frac{Kd-1}{T},
\end{equation}
according to the law of large numbers. The average SER is then given by
$\text{SER}(g)\bigr|_{g=G_0}$. That is, compared to a single user QPSK
transmission scheme, the performance for a multi-user system as we proposed
suffers an multiplicative SNR loss factor of $G_0$. Large $M$ is preferred for
better performance as expected. However, there is an effect of diminishing
returns based on \eqref{eq.G0}. Intuitively, a large $K$ is preferred for
increasing user detection probability. On the other hand, large $K$ causes
more SNR degradation. When $Kd$ approaches $MT$, symbol error probability is
expected to be large.

\section{ICBOMP Algorithm For Signal Recovery}\label{ICBOMP}

The previous analyses of normalized BOMP indicate that a smaller iteration
number $K$ can provide a lower SER. However, since $N_\text{a}$ is unknown at
the BS, $K$ must be large enough to accommodate a possibly large $N_\text{a}$.
As $Kd$ approaches $MT$, the performance of symbol demodulation may degrade
significantly. During the iterations, many blocks are expected to have only a
few demodulation errors, which can be corrected if an error control code is
employed. Error control coding is widely used for correcting demodulation
errors due to noise and interferences, and error detection codes such as
cyclic redundancy check (CRC) codes can be utilized to indicate whether the
decoded packets are error-free.

This motivates us to consider error control coding at the transmitters and an
improved ICBOMP algorithm for recovery. The ICBOMP algorithm makes use of
error correction and detection to carry out perfect interference cancellation
in each iteration of the normalized BOMP. Perfect interference cancellation
means that if some blocks of signals are correctly recovered before the
iterations have been finished, they are regarded as interference to the
following iterations and will be canceled to end their update. The idea of
interference cancellation can be found in \cite{ICanalysis1994}.

The ICBOMP is presented in Algorithm 2.

\subsection{ICBOMP algorithm}

In Algorithm 2, $\bbs_{\Lambda_k}^i$ is the $i$-th block of
$\bbs_{\Lambda_k}$ obtained by the LS reconstruction. Since some blocks of
signals may have been correctly recovered and then canceled by the
previous $k-1$ iterations, it has $1 \le i \le |\Lambda_{k-1}| +1 \le k$.
Vector $\tbs_{\Lambda_k}$ is the output of $\bbs_{\Lambda_k}$ after error
correction and detection. Let $\tLambda_k$ denote the index set for
error-free blocks at the $k$-th iteration.

\alglanguage{pseudocode}
\begin{algorithm}
\caption{ICBOMP} \label{algorithm2}
\begin{algorithmic}[1]
\State \textbf{Input: }the same as as normalized BOMP \State
\textbf{Initialization: }index set ${{\Lambda}_0} = \phi $,
residual signal ${\br_0} = \by$ \ForAll{$k \in \{1, 2, \cdots,
K\}$}
      \begin{enumerate}
      \item initialize a temporary index set: $\tLambda_k = \phi$
      \item calculate the $c_{j, k}$ for $j \in \left\{ [N]\backslash \{{\Lambda}_{k-1}~ \tLambda_{k-1}\cdots \tLambda_1\}\right\}$: $c_{j, k}={\| {{\bB_j^H}{\br_{k-1}}} \|_2^2/\|\bh_j\|_2^2}$
      \item find the index ${\lambda_k}$:
            $ {\lambda_k} = \arg \max_{j \in \left\{ [N]\backslash \{{\Lambda}_{k-1}~ \tLambda_{k-1}\cdots \tLambda_1\}\right\}} c_{j, k}$
      \item augment the index set:
       ${{\Lambda}_k}= {{\Lambda}_{k-1}} \cup \{ {\lambda_k}\}$

      \item update the signals by LS algorithm:
           ${\bbs_{{\Lambda}_k}} = \arg \min_{{\bs_0}}
              {\| {\by-\sqrt {{\rho_0}} {\bB_{\Lambda_k}}{\bs_0}}\|_2} $

     \item \textbf{for all} {$i \in \{1, 2, \cdots, |\Lambda_k|\}$} \textbf{do} the following error correction and detection:
        \begin{enumerate} [6a)]
         \item If $\bbs_{{\Lambda}_k}^i$ after error correction and detection is error-free,
         then $\tbs_{{\Lambda}_k}^i$ is the decoded version of $\bbs_{{\Lambda}_k}^i$;
         the original index of this $i$-th block should be added into $\tLambda_k$.
         \item If $\bbs_{{\Lambda}_k}^i$ after error correction and detection is not error-free, then
         $\tbs_{{\Lambda}_k}^i$ is the same as $\bbs_{{\Lambda}_k}^i$ without correction.
        \end{enumerate}
      \item \textbf{end for}

      \item perform perfect cancellation: $\by=\by-\sqrt {{\rho_0}} {\bB_{\tLambda_{k}}}{\tbs_{\tLambda_{k}}}$

      \item  update the index sets: $ \Lambda_k = \Lambda_k\backslash \{ \tLambda_k \} $

      \item update the residual signals:
           $ {\br_k} = \by- \sqrt {{\rho_0}}{\bB_{\Lambda_k}}{\tbs_{{\Lambda}_k}} $

      \end{enumerate}
\EndFor \State \textbf{Output: }$\tbs_K$ and the already correctly
recovered signal blocks
\end{algorithmic}
\end{algorithm}

It can be observed from the comparison of ICBOMP and normalized BOMP
algorithms that, their main difference occurs after signals have been
updated by the LS method. The ICBOMP will perform error correction and
detection for each block in $\bbs_{\Lambda_k}$. And when some blocks of
signals are decided to have been correctly recovered, ICBOMP will perform
the perfect cancellation. When perfect cancellation is allowed, ICBOMP can
offer the following benefits:
\begin{enumerate}
\item \emph{Higher user detection successful rate and signal demodulation
accuracy:}
      By error correction and detection, the accuracy of signal demodulation
      is improved, and this will reduce the error propagation which is the main drawback
      of the (normalized) BOMP. In turn, this is beneficial for the detection
      of remaining users and their signals' demodulation. Additionally,
      since error correction and detection are performed iteration by iteration for
      signals that are not previously error-free, the signals have a
      higher probability to be correctly recovered.

\item \emph{Lower computational cost:} When some blocks of signals have
been exactly recovered,
      The ICBOMP regards them as interference to the following iterations and no
      longer update these signals. For the LS update, more blocks indicate
      more computational cost. Therefore, the ICBOMP can reduce the computational cost.
\end{enumerate}

The benefit of ICBOMP is from the perfect interference cancellation after
error correction and detection. The price paid is a slightly decreased data
rate due to coding and a certain amount of additional calculations in
decoding. However, the reduced computational cost outweighs the additional
computational cost of decoding. As demonstrated by simulations, when the SNR
is not too small and under the same condition, by the BCH coding
scheme, the ICBOMP recovery runs much faster than the normalized BOMP. The
performance analysis by the ICBOMP recovery will require a lot of future work.

\section{Numerical Results}\label{simulations}

The simulation studies for verifying the proposed scheme are presented in
this section. In all simulations, QPSK is applied for data modulation.
Channel vectors are i.i.d.\ complex Gaussian with zero-mean and
unit-variance for each element. The $N_\text{a}$ active users are chosen
uniformly at random among all $N$ online users. We will choose the frame
length $T$ to be a multiple of the maximum length of short messages, $T=5d$
in all our presented simulations. All packets are assumed to have 100
nonredundant message-bearing symbols, i.e., 200 information bits. As for
precoding design, we simply generate a complex Gaussian matrix and then
normalize its columns to produce $\bP_n$, $n = 1,2, \cdots ,N$.

In our simulations, the SNR is defined as $E_s/N_0$, where $E_s$ is the symbol
energy and $N_0$ is the noise spectral density. The SER is computed as
follows: when a demodulated symbol of an active user is different from its
original symbol, we claim an error; if an active user is missed to detect,
then all its $d$ symbols are treated as erroneous. When an active user is
selected before algorithm finishes, we claim one successful detection, and the
number of all successful detections divided by $N_\text{a}$ is UDSR.

\subsection{Simulations with normalized BOMP}

In the first two experiments, we have the simulations of normalized BOMP
(labeled as NBOMP) when BS are equipped with 8 antennas.

\begin{figure}[ht]
\centering
\includegraphics[width=\figwidth]{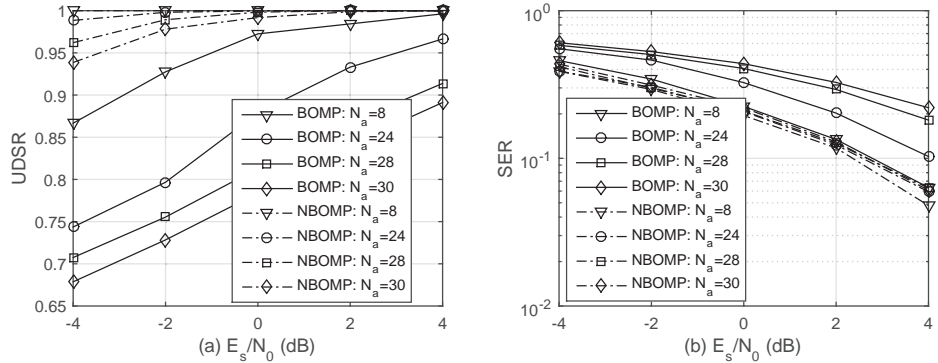}
\caption{performance of the normalized BOMP and the BOMP,
$(M,N,d,K)=(8,80,100,30)$}
\label{normalizedBOMP}
\end{figure}

\testcase \figref{normalizedBOMP} compares the performance of normalized
BOMP and BOMP algorithms. As we can see, for both algorithms, when the
number of active users increases, lower UDSR occurs. Obviously, the
normalization approach contributes to a much higher UDSR, which directly
results in a much lower SER, especially when $N_\text{a}$ is large like
$N_\text{a} \ge 24$. We also observe that, as long as UDSR exceeds a
certain value, for example, $\text{UDSR} > 85\%$, SERs of different numbers
of active users are very close.

\begin{figure}[ht]
\centering
\includegraphics[width=\figwidth]{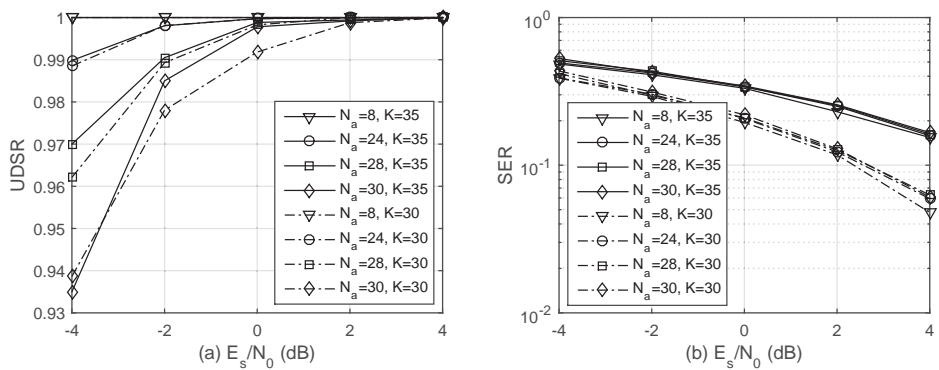}
\caption{performance of the normalized BOMP, $(M,N,d)=(8,80,100)$}
\label{M8bompK}
\end{figure}

\testcase Influences of iteration number $K$ on the performance of USDR and
SER are given in the \figref{M8bompK}. It can be drawn from the results that,
UDSR performance is improved with a larger $K$, see \figref{M8bompK} (a).
While \figref{M8bompK} (b) shows that, a larger $K$ will degrade the
performance of symbol demodulation, which is also indicated by our previous
analysis result of SER. Additionally, it is noted that, when $E_s/N_0=
4$dB, even 30 iterations are enough to select 30 active users.

\subsection{Simulations with ICBOMP}

In this part, we report results on ICBOMP algorithm. Channel coding
BCH(255,223) that can correct at most 4 erroneous bits and CRC of 16 bits
are acting as error correction and detection codes. With necessary zero
padding, we map 200 raw data bits and 16 CRC bits into 248 bits. Therefore,
it has $d=124$ and $T=5d=620$. BCH(255,223) still offers a high coding rate.
It should also be stated that BCH coding is not the only choice, we can
use other more efficient coding schemes. Compared with the previous simulations,
0.6dB reduction in $E_s/N_0$ is caused due to channel coding at the rate
$\frac{223}{255}$.

\begin{figure}[ht]
\centering
\includegraphics[width=\figwidth]{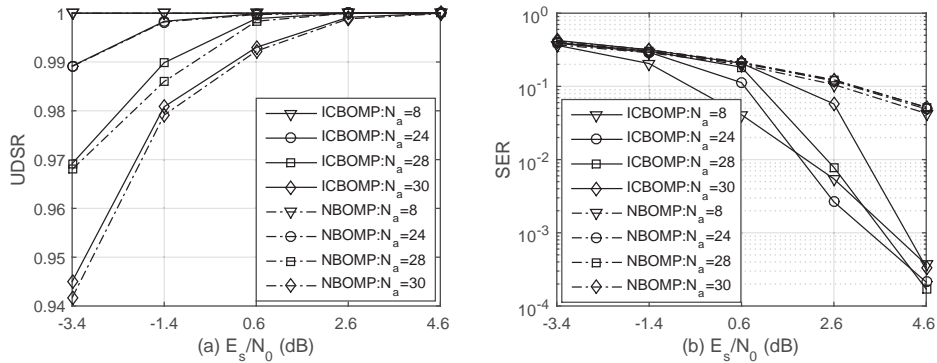}
\caption{performance of the ICBOMP and the normalized BOMP,
$(M,N,K,d)=(8,80,30,124)$}
\label{M8icbompbomp}
\end{figure}

\begin{figure}[ht]
\centering
\includegraphics[width=\figwidth]{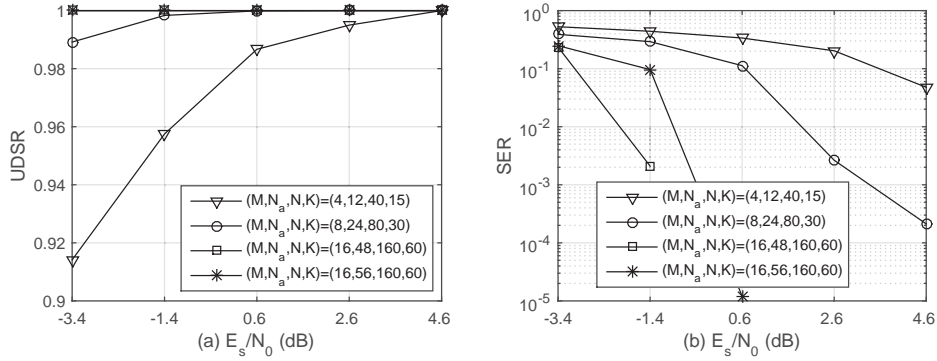}
\caption{performance of ICBOMP when BS is equipped with different numbers
of antennas, $d=124$}
\label{icbompM4816}
\end{figure}

\begin{figure}[ht]
\centering
\includegraphics[width=\figwidth]{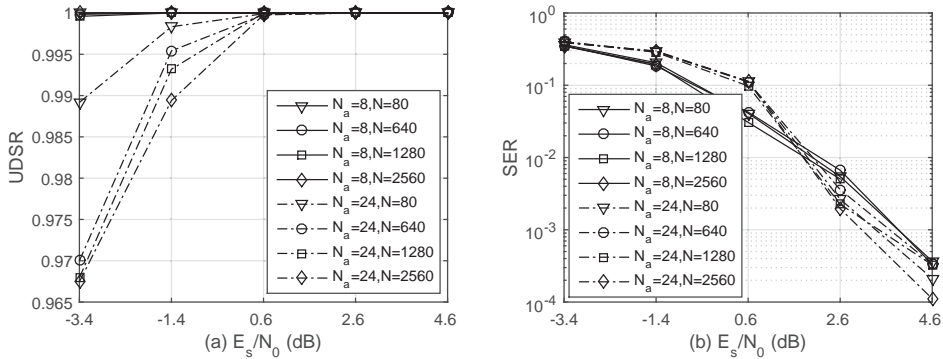}
\caption{performance of the ICBOMP when there are different numbers of online users,
$(M,K,d)=(8,30,124)$}
\label{M8NNa}
\end{figure}

\testcase \figref{M8icbompbomp} depicts the performance when ICBOMP is
exploited. Comparison with normalized BOMP is also made, where the same
error correction and detection are only performed at final iteration of the
normalized BOMP. As we can see, ICBOMP improves the performance, especially
in symbol demodulation when $E_s / N_0$ increases. In fact, whether
correction and detection coding is used when iterations finish almost makes
no difference for the normalized BOMP. However, for ICBOMP, since error
correction and detection are performed at each iteration, each block of
signals has a much higher probability to be correctly recovered.
Interestingly, we observe that, when almost all active users can be
selected, a larger $N_\text{a}$ offers a lower SER for a same $N$. This
result cannot be obviously deduced from analysis. It is related to the
order statistics of powers of channel vectors. Such phenomenon is
also expected to appear in the normalized BOMP when there are not too many
active users.

\testcase In \figref{icbompM4816} we have the simulation results when BS is
equipped with different numbers of antennas, respectively with 4, 8, and 16
antennas. To have a fair comparison, we set different parameters to
guarantee the ratios of $(M : N : N_\text{a} : K)$ at a same level, and
to show the benefits of $M=16$, one additional result of larger $N_\text{a}
/ M$ $(i.e., N_\text{a} = 56)$ is presented. The results show that, more
antennas at BS can offer remarkable benefits, both in user detection and
symbol demodulation, and these benefits are much more than linear with
the antenna number $M$.

\testcase It is desirable that a system can accommodate many online users.
When BS is equipped with 8 antennas, the UDSRs and SERs of different
numbers of online users are depicted in the (a) and (b) of \figref{M8NNa}. As
we can see, although a larger number of online users $N$ decreases the UDSR, the
degradation from $N=80$ to $N=2560$ is not very obvious for a same
$N_\text{a}$. When $N_\text{a}$ is not large, like $N_\text{a}=8$ in our
simulation, such degradation can be neglected. SERs of different numbers of
online users are rather close, as shown in the \figref{M8NNa} (b). It means
that our proposed scheme with ICBOMP receiver for user detection and symbol
demodulation is applicable for a system with many online users. As the same
in \figref{M8icbompbomp}, when all active users can be selected in the higher
SNR regime, SER of larger $N_\text{a}$ is lower.

Combination of the results in \figref{icbompM4816} and \figref{M8NNa} gives us
the confidence that, when BS is equipped with many antennas, e.g., in
massive MIMO systems, more online users and active users
are allowed to exist. This makes the proposed scheme especially suitable
for the future 5G systems to support massive connections.

Different frame lengths also offer different performance. As a conclusion,
since $T$ acts as block length in channel coding, larger $T$ offers better
performance in the relatively high SNR regime.

\subsection{Computational cost reduction by ICBOMP}

\begin{figure}[ht]
\centering
\includegraphics[width=\figwidth]{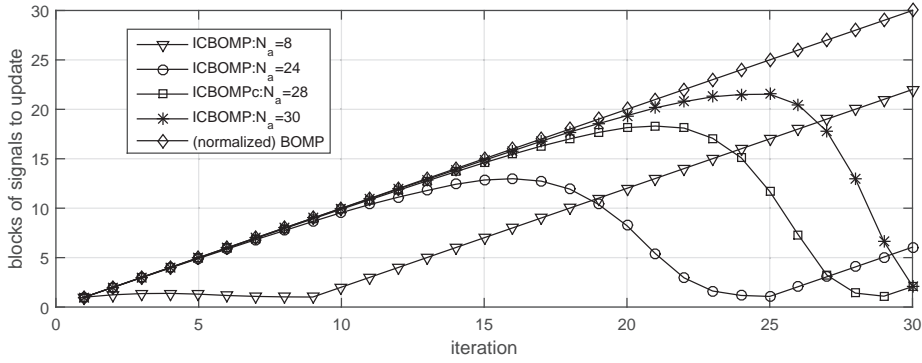}
\caption{computational cost reduction by the ICBOMP,
$(M,N,K,d)=(8,80,30,124)$}
\label{icbompanalysis}
\end{figure}

\testcase In this part, we cite one example to demonstrate the benefit of
ICBOMP in computational cost reduction. The results are shown in the
\figref{icbompanalysis} and they are obtained when $E_s/N_0 = 4.6$dB. The
ordinate is the number of blocks for signal update at each iteration, which
is directly related to the computational cost. The result of the
(normalized) BOMP is also presented. It shows that under the given
condition, some blocks of signals can be correctly recovered before
the iterations finish. For example, for $N_\text{a}=30$ under the given
condition, the maximum number of blocks of signals to update is about 22,
and after that, the number quickly decreases. However, for the (normalized)
BOMP, since no blocks will be canceled, its number of blocks for update
increases linearly with the iteration number. The benefit of ICBOMP in
reducing the computational cost is therefore demonstrated.

\subsection{Results of the approximate analysis}

\begin{figure}[ht]
\centering
\includegraphics[width=\figwidth]{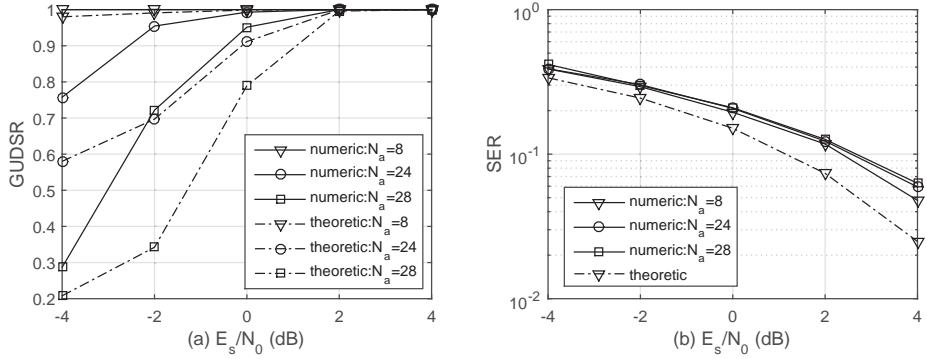}
\caption{comparison of simulation and analysis of the normalized BOMP when
$(M,N,K,d)=(8,80,30,100)$}
\label{bompanalysis}
\end{figure}

\testcase In this part, we show two examples to compare our analyses with
the simulations, including the user detection in \figref{bompanalysis} (a)
and the symbol demodulation in \figref{bompanalysis} (b). User detection
performance is quantified using GUDSR.

It can be observed from the \figref{bompanalysis} (a) that, our lower bound
analysis about GUDSR of the normalized BOMP has certain gap to the simulation
results when $E_s/N_0$ is below a certain level. It is because in this
regime, except the most likely case we considered for successful user
detection, other possible cases also contribute a high probability for the
GUDSR. As $E_s/N_0$ increases, since the cited case plays a more and more
dominant role in the successful user detection, gradually, this gap narrows
and the analysis presents a tighter lower bound. On the other hand, our
approximate analysis about the SER shows stable performance, about $1.5$dB better
than the simulation result. Since our SER analysis is based on the case that all active
users have been successfully identified, it remains the same for different numbers of the
active users. That is why only one analytical curve is presented.

\subsection{Probabilistic transmission}

\begin{figure}[ht]
\centering
\includegraphics[width=\figwidth]{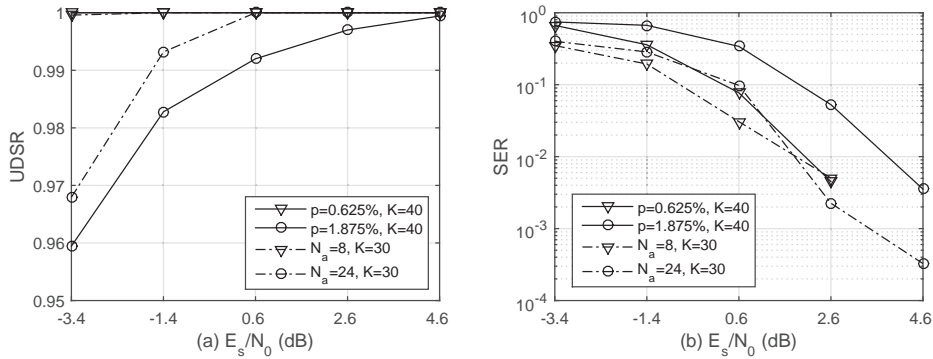}
\caption{performance of the ICBOMP when each user transmits with a certain
probability, $(M,N)=(8,1280)$}
\label{probability}
\end{figure}

\testcase In order to exactly test the influences of different parameters
such as $M$ and $N$ on the performance, we have assumed a constant $N_\text{a}$ for each set
of parameters. However, random transmission with certain active probability
$p$ makes more sense for certain applications. In this part, we have
simulations for random transmission where each user sends a small packet
according to a probability $p$. Since $N=1280$,
probabilities $p=0.625\%, 1.875\%$ are expected to respectively produce
$N_\text{a}=8, 24$ active users. We set $K=40$ for this probabilistic
transmission case. See \figref {probability} for the results.

Under the given condition, almost all results of different active
probabilities show performance degradation when compared with the results
of given $Np$ active users, especially when $p$ is large. When $p=0.625\%$,
this degradation is not severe and SER degradation mainly comes from a
larger iteration number $K$, which has been previously demonstrated.
Such slightly performance degradation is also the case when $p=1.25\%$,
where $N_\text{a}=16$ is expected. When $p=1.875\%$, if we use Gaussian
distribution to estimate the number of active users, then $N_\text{a} $
can be larger than 47 with a probability $15.85\%$. Since the number of
active users can be even larger than $K$, it will certainly make some
active users missed to detect. This missed alarm greatly harms the
performance of symbol demodulation. In fact, the ICBOMP receiver can handle
this practical problem to some extent. When signals of one active user have
been correctly recovered, set $K=K+1$. Then, there
are more iterations for user detection. Fundamentally, the most effective
technique to solve this problem is to equip the BS with more antennas to
support more online users and active users.

\section{Discussions}\label{Discussion}

In this section, we discuss three issues that are related to our proposed
scheme.

\subsection{Comparison with known schemes}

Our proposed scheme is different from the existing schemes that are based on
the orthogonal access patterns, e.g., time-division multiple access (TDMA),
CDMA and single carrier orthogonal frequency division multiplexing (SC-FDMA).
TDMA is mostly for the 2G wireless systems, CDMA is deployed in the 3G
wireless systems and SC-FDMA is proposed for the uplink in the LTE systems
\cite{myung2006single}. To meet the requirements of future 5G wireless
network, e.g., offering massive connectivity, several non-orthogonal multiple
access schemes have been proposed
\cite{SCMA2013,SCMAcodebook2014,NOMA2015,MUSA2015}, including sparse code
multiple access (SCMA) \cite{SCMA2013,SCMAcodebook2014}. Since the number of
resources are strictly limited for orthogonal multiple access schemes, the
number of users that can be served is limited. Furthermore, resource
scheduling is necessary. In the case of burst transmissions of many small
packets, dynamic resource scheduling leads to significant signaling overhead.
In contrast, our scheme requires no such dynamic scheduling and hence improves
the resource utilization efficiency. During one frame interval, since
different precoding codes are used for different symbols and even the same
symbol can be assigned different precoding codes at different times, the
interferences among symbols are averaged. This is favorable for improving the
recovery performance. Our system also has robust performance in detecting user
activities by incorporating error correction and detection into the recovery
algorithm. The comparisons among TDMA, CDMA, SC-FDMA, SCMA and our scheme are
summarized in Table~\ref{compschemes}. For the future 5G network, our proposed
non-orthogonal many access scheme is promising to address the challenges
produced by the transmissions of small packets.

\begin{table}[htb]
\caption{Comparisons of different schemes}\label{compschemes} \centering
\begin{tabular}{|c|c|c|c|c|}
\hline
            & TDMA/SC-FDMA  & CDMA & SCMA & Our Scheme   \\
\hline Coding Level & $\diagup$ & symbol & bit & frame \\ \hline
Codebook Design & $\diagup$ & independent of CSI & dependent on CSI &
independent of CSI \\ \hline Resource Scheduling & necessary & necessary &
necessary & unnecessary \\ \hline Massive Connectivity & & & & $\alpha
\frac{MT}{d}$ (e.g., $\alpha >
0.7$ \\ support ($N_\text{a}$ per resource &limited, at most 1 &
limited, at most $\frac{T}{d}$ & unclear when $M > 2$ &
when $T=5d$, $M =16$ \\ block with $T$ units) & & & & in \figref{icbompM4816}) \\ \hline
\end{tabular}
\end{table}

\subsection{Message segmentation for practical transmission}

Although the length of small packet is short, block-sparse signal recovery by
ICBOMP still requires heavy computations, especially when $d$ is not very
small and there are many receiving antennas. To reduce such cost, messages of
small packet can be segmented into shorter parts further. These segments can
be successively transmitted and signal recovery is successively performed for
them. For different segments of one small packet, we can even pre-assign
different precoding matrices to them. There are several benefits of using such
segmented transmission. Firstly, it can reduce the computational cost for
signal recovery. Secondly, it can alleviate the length requirement for constant duration
of the block-fading channel. In scenario where channel varies in short period, pilots
can be inserted between two segments for the CSI estimation or renewal.
Additionally, it can also avoid having highly correlated blocks in $\bB$ for
large duration. Highly correlated channel vectors can be essentially
decorrelated at the BS when different precoding matrices for different
segmentation periods are used.

\subsection{Sparsity level estimation}

Since the receiver does not know the number of active users $N_\text{a}$, a
large number of iterations are set for the recovery algorithms. When
$N_\text{a}$ is small and user detection can be successful with less number of
iterations. However, unnecessary iterations not only increase the computational
cost but also weaken the ability for an exact signal recovery \cite{re40}. Such
phenomenon has also been observed in our previous analysis and simulations.
Our numerical simulation indicate that, when $E_s/N_0 \ge 2.6$dB and with the
conditions $M=8$, $N=2560$, $T=5d=620$, 24 iterations for ICBOMP will
correctly identify all the 24 active users with a probability exceeding
99.99\%. In this case, technique for sparsity level (i.e., $N_\text{a}$)
estimation will provide benefits. For example, work to estimate the random
sparsity level can be found in \cite{re40,QiuKun2012}. However for
block-sparsity scenario, more effective technique will be useful.

\section{Conclusions}\label{Conclusion}

In this paper, we proposed an uplink many access scheme for the services of
small packet. The proposed scheme combines the techniques of block
precoding at the transmitters and sparsity-aware recovery at the receiver.
The proposed non-orthogonal transmission scheme is applicable to future
wireless communication systems with many receive antennas. The reason is
that in the future communications, small packets play a more and more important
role due to the rapid development of the mobile Internet. The overall
throughput of such systems is currently hampered by small packets because
of the heavy signaling overhead related to transmitting them. Designed to
solve this problem, our scheme can greatly reduce the signaling overhead
and therefore guarantees a high throughput. Besides, future communication
systems are supposed to have massive antennas at receiver, which enables
the proposed scheme to offer massive connections for the frequent and
random transmissions of small packets.

\appendices
\section{Brief Derivations of \eqref{mean00} and \eqref{mean01}}

Due to the space limitation, we only present the important steps of the
following derivations. The derivations are based on the previously made
approximations.

At the $k$-th iteration, the residual signal vector is given by
\eqref{residualt0c}. For $i\neq j$, symbol vector $\bs_i$ is independent of
$\bs_j$ and channel vector $\bh_i$ is independent of $\bh_j$. When $j
\notin I$, we have that
\begin{equation} \label{important00}
\begin{array}{rcl}
\mu_{0,k,j}
&=& \mathbb E \left\{ \frac{\| \bB_j^H \br_{k-1} \|_2^2}{\bh_j^H \bh_j}\right\}
= \mathbb E \left\{ \frac{\br_{k-1}^H \bB_j \bB_j^H \br_{k-1} }{\bh_j^H \bh_j}\right\} \\
&=& \mathbb E \left\{ \frac{\rho_0 \sum_{i \in I \backslash {\Lambda_{k-1}}} \bs_i^H \bB_i^H \bB_j \bB_j^H \bB_i \bs_i + \tbz_k^H \bB_j \bB_j^H \tbz_k }{\bh_j^H \bh_j} \right\} \\
&\approx& \rho_0 (N_\text{a}-s_{k-1}) \mathbb E \left\{ \bs_i^H \bP_i^H \bP_j
\bP_j^H \bP_i \bs_i \right\} + \sigma_k^2 \trace \left\{ \mathbb E \left\{ \bP_j^H \bP_j \right\}\right\} \\
&\approx& \rho_0 (N_\text{a}-s_{k-1}) \frac{d^2}{T}  + d \sigma_k^2
\end{array}
\end{equation}
where the identity $\bB_i^H \bB_j=\left( {\bP_i^H \otimes \bh_i^H} \right)
\left( {\bP_j \otimes \bh_j} \right) = \left(\bP_i^H \bP_j\right) \otimes
\left(\bh_i^H \bh_j\right) =\left(\bP_i^H \bP_j\right) \left(\bh_i^H
\bh_j\right)$, the fact $\mathbb E \left\{ \tbz_k^H \bB_j \bB_j^H \tbz_k
\right\} = \sigma_k^2 \trace\left\{ \mathbb E \left\{ \left(\bP_j^H
\bP_j\right) \left(\bh_j^H \bh_j\right)\right\}\right\}$ and approximation
$\mathbb E \left\{ \frac{\bh_j \bh_j^H}{\bh_j^H \bh_j} \right\} \approx
\frac{I_M}{M}$ have been used for the above derivations. $\tbz_k$ is
assumed to be vector with i.i.d.\ variables. By the assumption that
entries of precoding matrices are 0-mean and $\frac{1}{T}$-variance i.i.d.
Gaussian variables, $\mathbb E \left\{ \bs_i^H \bP_i^H \bP_j \bP_j^H \bP_i \bs_i
\right\} = \mathbb E \left\{ \sum_{m=1}^{d}\sum_{n=1}^{d}\sum_{r=1}^{d}
\bs_{i,m}^H \bP_{i:m}^H \bP_{j:r} \bP_{j:r}^H \bP_{i:n} \bs_{i,n}\right\}
=\frac{d^2}{T}$, where $\bs_{i,n}$ and $\bP_{i:n}$ respectively denote the
$n$-th element of $\bs_i$ and $n$-th column of $\bP_i$.

When $j\in I$, we have
\begin{equation} \label{important01}
\begin{array}{rcl}
\mu_{1,k,j}
&=& \mathbb E \left\{ \frac{\| \bB_j^H \br_{k-1} \|_2^2}{\bh_j^H \bh_j}\right\} \\
&=& \mathbb E \left\{ \frac{\rho_0 \bs_j^H \bB_j^H \bB_j \bB_j^H \bB_j \bs_j + \rho_0 \sum_{i \in I \backslash \left\{{\Lambda_{k-1}}, j \right\}} \bs_i^H \bB_i^H \bB_j \bB_j^H \bB_i \bs_i + \tbz_k^H \bB_j \bB_j^H \tbz_k}{\bh_j^H \bh_j}\right\}  \\
&\approx & \rho_0 \mathbb E \left\{ \left[\bs_j^H\left( \bP_j^H \bP_j \right)^2\bs_j\right]\left(\bh_j^H \bh_j\right) \right\} + \rho_0 (N_\text{a}-s_{k-1}-1) \frac{d^2}{T} + d \sigma_k^2 \\
&\approx & \rho_0 \frac{d}{T} (d+T-1) \mathbb{E} \left\{\bh_j^H \bh_j \right\} + \rho_0 (N_\text{a}-s_{k-1}-1) \frac{d^2}{T}  + d \sigma_k^2
\end{array}
\end{equation}
where the result of $\mu_{1,k,j}$ in \eqref{important00} is used, and
$\bs_j^H\left( \bP_j^H \bP_j \right)^2\bs_j$ is also expanded for the above
evaluation.

\section{Brief Derivations of \eqref{mean0}-\eqref{var1}}

The derivations are also based on the previously made approximations.

It should be noted that, $\mathbb E \left\{ \frac{\bh_i^H \bh_j \bh_j^H
\bh_i}{\bh_j^H \bh_j} \right\} \approx \frac{I_M}{M} \mathbb E \left\{
\bh_i^H \bh_i \right\}$, which is $I_M$ when $\bh_i^H \bh_i$ is not
ordered. In the derivations of Appendix A, by evaluating the $\mathbb E
\left\{ \bh_i^H \bh_i \right\}$ for different active users with order
statistics, we can respectively obtain $\mu_{0,k}$ in \eqref{mean0} and
$\mu_{1,k}$ in \eqref{mean1}.

For $j\notin I$, by the independence of $\bB_j$ and $\br_{k-1}$, it has that
$\mathbb E \left\{ \frac{\br_{k-1}^H\bB_j \bB_j^H \br_{k-1} }{\bh_j^H
\bh_j}\right\} = d \sigma_{\br_{k-1}}^2$. By this, we can assume that each
element of $\frac{\bB_j^H \br_{k-1}}{\sqrt{\bh_j^H \bh_j}}$ is a complex
Gaussian variable with a mean 0 and a variance $\sigma_{\br_{k-1}}^2$. The
variable $\frac{ \left(\br_{k-1}^H \bB_j \bB_j^H \br_{k-1}\right)}{ \bh_j^H
\bh_j}$ then follows a chi-squared distribution with $2d$ degrees of
freedom. Therefore, we can have the following evaluation
\begin{equation} \label{OrderJnotinI}
\begin{array}{rcl}
\mathbb E \left\{ \left(\frac{\| \bB_j^H \br_{k-1} \|_2^2}{\bh_j^H \bh_j}\right)^2\right\}=\mathbb E \left\{\left( \frac{ \br_{k-1}^H \bB_j \bB_j^H \br_{k-1} } {\bh_j^H \bh_j }\right)^2 \right\}
\approx d(d+1)\sigma_{\br_{k-1}}^4
\end{array}
\end{equation}
which gives the result in \eqref{var0}.

Define $\tbr_{k}: =\br_{k-1}-\sqrt{\rho_0} \bB_k \bs_k$, each of whose zero-mean
elements has a variance given by \eqref{residulcovariance}. Then for $j=k\in
I$, we have
\begin{equation} \label{important11}
\begin{array}{rcl}
\mathbb E \left\{ \left(\frac{\| \bB_j^H \br_{k-1} \|_2^2}{\bh_j^H \bh_j}\right)^2\right\}
&=& \mathbb E \left\{ \frac{\left( \rho_0 \bs_j^H \bB_j^H \bB_j \bB_j^H \bB_j \bs_j + 2 \sqrt{\rho_0} Re \left\{\bs_j^H \bB_j^H \bB_j \bB_j^H \tbr_k \right\} + \tbr_k^H \bB_j \bB_j^H \tbr_k\right)^2 }{\bh_j^H \bh_j \bh_j^H \bh_j} \right\} \\
&=& \rho_0^2 \mathbb E \left\{\left( \frac{ \bs_j^H \bB_j^H \bB_j \bB_j^H \bB_j \bs_j }{\bh_j^H \bh_j}\right)^2 \right\} + 2 \rho_0 \mathbb E \left\{ \frac{\left| \bs_j^H \bB_j^H \bB_j \bB_j^H \tbr_k \right|^2}{\bh_j^H \bh_j \bh_j^H \bh_j} \right\}  \\
&~& + \mathbb E \left\{\left(  \frac{ \tbr_k^H \bB_j \bB_j^H \tbr_k} {\bh_j^H \bh_j }\right)^2 \right\} + 2 \rho_0 \mathbb E \left\{\frac{ \left(\bs_j^H \bB_j^H \bB_j \bB_j^H \bB_j \bs_j\right) \left( \tbr_k^H \bB_j \bB_j^H \tbr_k \right)} {\bh_j^H \bh_j \bh_j^H \bh_j}\right\}.
\end{array}
\end{equation}

With the same approach to evaluate $\mathbb E \left\{ \frac{\bs_i^H \bB_i^H
\bB_j \bB_j^H \bB_i \bs_i}{\bh_j^H \bh_j} \right\}$ in \eqref{important00},
we have
\begin{equation}
\begin{array}{rcl} \label{important12}
\mathbb E \left\{\left( \frac{ \bs_j^H \bB_j^H \bB_j \bB_j^H \bB_j \bs_j }{\bh_j^H \bh_j}\right)^2 \right\}
\approx \frac{\rho_0^2}{T^2} (d^2+d)(d+T-1)^2 \mathbb{E} \left\{\left(\bh_{j}^H \bh_{j} \right)^2 \right\}.
\end{array}
\end{equation}

Furthermore, we have the following result
\begin{equation} \label{important13a}
\begin{array}{rcl}
\mathbb E \left\{\frac{ \left(\bs_j^H \bB_j^H \bB_j \bB_j^H \bB_j \bs_j\right) \left( \tbr_k^H \bB_j \bB_j^H \tbr_k \right)} {\bh_j^H \bh_j \bh_j^H \bh_j}\right\}
&=& \mathbb E \left\{\frac{ \left(\bs_j^H \bB_j^H \bB_j \bB_j^H \bB_j
\bs_j\right) \trace \left\{\bB_j^H \tbr_k  \tbr_k^H \bB_j \right\}} {\bh_j^H \bh_j \bh_j^H \bh_j}\right\}  \\
&\approx & \sigma_{\tbr_k}^2 \mathbb E \left\{\frac{ \left(\bs_j^H \bP_j^H
\bP_j \bP_j^H \bP_j \bs_j\right) \left( \bh_j^H \bh_j \right)^2 \trace \left\{\bP_j^H \bP_j \right\} \left( \bh_j^H \bh_j\right)} {\bh_j^H \bh_j \bh_j^H \bh_j}\right\} \\
&=& d \sigma_{\tbr_k}^2 \mathbb E \left\{ \bs_j^H \bP_j^H \bP_j \bP_j^H \bP_j \bs_j \right\} \mathbb E \left\{ \bh_j^H \bh_j \right\} \\
&\approx & \sigma_{\tbr_k}^2 \frac{d^2}{T} (d+T-1) \mathbb E \left\{ \bh_j^H \bh_j \right\}.
\end{array}
\end{equation}

Similar to have \eqref{OrderJnotinI}, we obtain
\begin{equation} \label{important13b}
\begin{array}{rcl}
\mathbb E \left\{\left(  \frac{ \tbr_k^H \bB_j \bB_j^H \tbr_k} {\bh_j^H \bh_j }\right)^2 \right\}
\approx d(d+1)\sigma_{\tbr_k}^4.
\end{array}
\end{equation}

Besides, the following derivation holds for $j\in I$
\begin{equation} \label{important20}
\begin{array}{rcl}
\mathbb E \left\{ \frac{\left| \bs_j^H \bB_j^H \bB_j \bB_j^H \tbr_k \right|^2}{\bh_j^H \bh_j \bh_j^H \bh_j} \right\}
&=& \mathbb E \left\{ \frac{\bs_j^H \bB_j^H \bB_j \bB_j^H \tbr_k \tbr_k^H \bB_j \bB_j^H \bB_j \bs_j}{\bh_j^H \bh_j \bh_j^H \bh_j} \right\} \\
&\approx & \sigma_{\tbr_k}^2 \mathbb E \left\{ \left[\bs_j^H\left( \bP_j^H \bP_j \right)^3\bs_j\right]\left(\bh_j^H \bh_j\right) \right\} \\
&=& d \sigma_{\tbr_k}^2 ~ \Bar {\lambda}^3 \mathbb{E} \left\{\bh_{j}^H \bh_{j} \right\}
\end{array}
\end{equation}
where $\Bar {\lambda}^3$ is the expected eigenvalue of $\left( \bP_j^H
\bP_j \right)^3$.

Define $\beta:=\frac{d}{T} \in (0, 1)$, by the asymptotic result of
Theorem 2.35 of random matrix theory in \cite{RMT2004}, which is also
a good approximation for reasonably small-scale matrix dimensions, the
eigenvalues of $\left( \bP_j^H \bP_j \right)$ have an empirical distribution
\begin{equation} \label{rmtPDF}
f_\beta(x)=\frac{1}{2 \pi \beta x} \sqrt{(x-a)(b-x)}
\end{equation}
for $x\in [a, b]$, where $a=\left(1-\sqrt \beta\right)^2$ and $b=\left(1+\sqrt \beta\right)^2$.

Then $\Bar {\lambda}^3$ can be approximately evaluated as
\begin{equation} \label{important21}
\begin{array}{rcl}
\Bar {\lambda}^3 \approx \int_{\left(1-\sqrt \beta\right)^2}^{\left(1+\sqrt \beta\right)^2}  {\frac{x^3}{2 \pi \beta x} \sqrt{\left[x-\left(1-\sqrt \beta\right)^2\right]\left[\left(1+\sqrt \beta\right)^2-x\right]}} dx = \frac{1}{8} \left(3+\frac{9d}{T}-\frac{2d^2}{T^2}\right).
\end{array}
\end{equation}

Similarly, by the eigenvalue approach, $\mathbb E \left\{
\left[\bs_j^H\left( \bP_j^H \bP_j \right)^2\bs_j\right]
\right\}\approx\frac{d}{T} (d+T) $, which is almost the same as the
achieved $\frac{d}{T} (d+T-1)$ in \eqref{important13a}.

Finally, by substituting the results from \eqref{important12} to
\eqref{important21} into \eqref{important11} and by replacing $j$ with $k$
(indicating the $k$-th active user), we can have the derived result in
\eqref{var1}.

It should be noted that, when $k=N_\text{a}$, the $\mathbb E \left\{
\left(\frac{\| \bB_j^H \br_{k-1} \|_2^2} {\bh_j^H \bh_j}\right)^2\right\}$
for $j\notin I$ can be more precisely evaluated. As the same to have the
final result of \eqref{important11}, the more precise evaluation is
obtained by the expansion approach. It produces the following result
\begin{equation} \label{var002}
\begin{array}{rcl}
\mathbb E \left\{ \left(\frac{\| \bB_j^H \br_{N_\text{a}-1} \|_2^2}{\bh_j^H \bh_j}\right)^2\right\}
&\approx &  \frac{\rho_0^2d}{M^2 T^3} \left[ 4(dT+1)+(d^2-d) (dT+d+T) \right] \mathbb{E} \left\{\left(\bh_{N_\text{a}}^H \bh_{N_\text{a}} \right)^2 \right\} \\
&~& + 2 \rho_0 \frac{d^2}{M T^2} (dT+d+T-1)\mathbb{E} \left\{\bh_{N_\text{a}}^H \bh_{N_\text{a}} \right\} \sigma_{N_\text{a}}^2 + d(d+1)\sigma_{N_\text{a}}^4.
\end{array}
\end{equation}

\ifCLASSOPTIONcaptionsoff
  \newpage
\fi

\end{document}